\documentclass[aps,pra,floatfix,amsmath,amssymb,showpacs,showkeys,twocolumn]{revtex4}
\usepackage[caption=false]{subfig}
\usepackage{graphicx,bm,color}
\usepackage{amsfonts}

\begin{document}

\frenchspacing

\title{Entanglement propagation of a quantum optical vortex state}

\begin{abstract}
We study the entanglement evolution of a quantum optical vortex state propagating through coupled lossless waveguides. We consider states generated by coupling two squeezed modes using a sequence of beam splitters and also by subtracting photons from the signal in spontaneous parametric down conversion. We reconstruct the Wigner function at a later time to study the correlation and quantify the entanglement after propagation using \emph{logarithmic negativity}.
\end{abstract}

\author{Anindya Banerji$^{1,2}$\footnote{abanerji09@gmail.com}}
\author{Ravindra Pratap Singh$^{3}$\footnote{rpsingh@prl.res.in}}
\author{Dhruba Banerjee$^{1}$\footnote{dhruba.iacs@gmail.com}}
\author{Abir Bandyopadhyay$^{2}$\footnote{abir@hetc.ac.in}}
\affiliation{$^{1}$Department of Physics, Jadavpur University, Kolkata 700032, India}
\affiliation{$^{2}$Hooghly Engineering and Technology College, Hooghly 712103, India}
\affiliation{$^{3}$Physical Research Laboratory, Ahmedabad 380009, India}

\date{\today}


\maketitle

\section{Introduction}

Recently a lot of work has been done to study states with non$-$Gaussian quadrature space distribution, which find numerous applications in quantum information processing. Experiments have been conducted to produce non$-$Gaussian states using techniques like homodyne detection from a single mode squeezed state of light \cite{Wenger,Kim}. Photon subtraction/addition from a single mode squeezed state also gives rise to non$-$Gaussian states with associated Wigner function showing negative regions \cite{Asoka}. The negativity of the Wigner function has been studied as a measure of nonclassicality for these states \cite{Kenfack}. These states are important since they find a variety of useful applications in quantum computation \cite{Bartlet}, entanglement distillation \cite{Eisert,Giedke} and loophole free tests of Bell's inequality \cite{Nha}.
Another class of non$-$Gaussian states with interesting nonclassical features and a negative Wigner function are the quantum optical vortex states, introduced in \cite{GSA97} and studied in some detail in \cite{jbvortex,Abir PLA,Abir OptComm,gsa_njp,ABanerji,anindyaarxiv}. These are states with topological defects in the phase space and have a vortex structure in the quadrature space. Such states can be generated from two mode squeezed vacuum under a linear transformation belonging to the SU(2) group with certain restrictions \cite{jbvortex}. These states have been realized in the laboratory using photon subtraction \cite{Parigi}. It has been pointed out that photon subtraction/addition leads to enhancement of entanglement \cite{CNB}. In this article we deal with vortex states arising from photon subtraction from one of the modes of a two mode squeezed vacuum. The order of the vortex is determined from the number of photons subtracted \cite{anindyaarxiv}. Interestingly, this state carries OAM, given by $m\hbar$, when $m$ photons are subtracted.\\
We also consider vortex states produced by mixing two squeezed modes using a beam spliiter (BS) or a dual channel directional coupler (DCDC) \cite{Abir OptComm}. It should be mentioned that a similar state can be generated by using a $\Lambda$ type three level atom with counter rotating photons having circular polarization and performing a conditional measurement \cite{GSA97}. Entanglement being a fundamental resource in quantum information processing, it is interesting to study states with enhanced entanglement from a task oriented point of view. It has been shown that vortex states carry more entanglement compared to the Gaussian states from which they are generated and the entanglement carried can be controlled by altering the squeezing parameter or the ratio of mixing of the two input modes in a beam splitter \cite{ABanerji}.\\
In this article we study the propagation of entanglement of the generalized vortex state using coupled lossless waveguides. The importance of coupled waveguides lie in their efficiency to manipulate the flow of light \cite{Peschel,Pertsch,Morandotti,Lederer,Longhi,Longhi2,Bromberg,ARai}. They have been used extensively to implement quantum random walk problems \cite{Perets} which finds important applications in quantum computation and quantum algorithms. Coupled waveguides have been successfully used to implement a CNOT gate on a silica chip \cite{Politi}. Given all these developments, it is important to study how such a system affects the entanglement present in the light moving through it. We use \emph{Logarithmic Negativity} as a quantitative measure to study the behavior of entanglement on propagation through coupled waveguides. We also calculate and study the associated Wigner function. Time evolution of the Wigner function has been a difficult problem due to its negative values \cite{Wong}. We use a numerical approach to evolve the Wigner function as a function of the phase space coordinates and use it to study the quantum correlations between the two modes at a later time.\\
The article is organized as follows. In the next section we briefly introduce the model used for vortex evolution. In section \ref{sec:PS}, we solve for the time evolution of the photon subtracted vortex states. We present the equation and an interpretation and explanation for the results in the same section. In section \ref{sec:BS} we discuss the time evolution of the vortex state generated by coupling two squeezed modes using a series of beam splitters and explain the results obtained therein. We also construct the Wigner functions for the respective states at a later time. In section \ref{sec:Ent}, we present an explicit approach for studying the entanglement and its variation with time for both the states using \emph{logarithmic negativity}.\\

\section{Generalized vortex under time evolution}
The model that we consider here consists of two single mode coupled waveguides. The Hamiltonian for this system \cite{Lai} can be written as follows
\begin{equation}
\label{Hamiltonian}
H=\hbar\omega\left(a^{\dagger}a+b^{\dagger}b\right)+\hbar C(a^{\dagger}b+b^{\dagger}a)
\end{equation}
where \emph{a} and \emph{b} are the regular bosonic mode operators for the two single mode waveguides. The first two terms correspond to the free energy while the next two terms take into account the evanescent coupling between the two waveguides with \emph{C} as the coupling strength. Since we consider lossless propagation over short time intervals, the Heisenberg equations of motion can be used to study the time evolution of the bosonic field operators, \emph{a} and \emph{b} for the two modes \cite{Rai}. The time dependence of these operators are then given as
\begin{eqnarray}
\label{modeoperatorstimedep}
a(t)=a(0)\cos(Ct)-\text{i}b(0)\sin(Ct) \nonumber \\
b(t)=b(0)\cos(Ct)-\text{i}a(0)\sin(Ct)
\end{eqnarray}
We use Eq. \ref{modeoperatorstimedep} to study the time evolution of the generalized vortex states. We consider two different input states. One of them is generated by subtracting \emph{k} photons from one of the modes of a two mode squeezed vacuum which we study in section \ref{sec:PS}. The other one can be generated by using \emph{k} beam splitters to couple two squeezed mode vacuum states which we study in section \ref{sec:BS}. The difference between these two states is that the former is already entangled before the process of photon subtraction while the latter gets correlated after being coupled by the beam splitters.\\

\subsection{Photon Subtraction}
\label{sec:PS}
\begin{figure*}
\centering
\subfloat[]{
\includegraphics[scale=0.2]{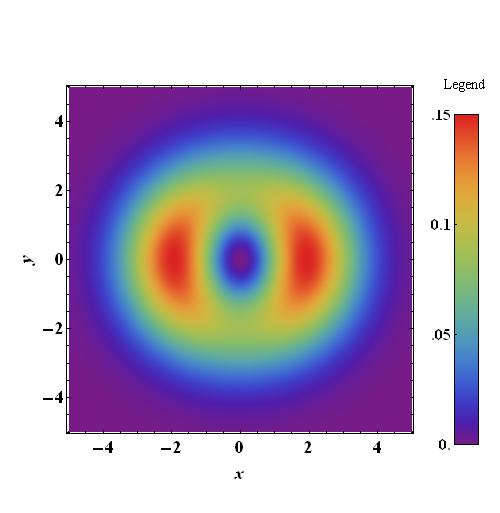}
\label{fig:label-e}}
\qquad
\subfloat[]{
\includegraphics[scale=0.2]{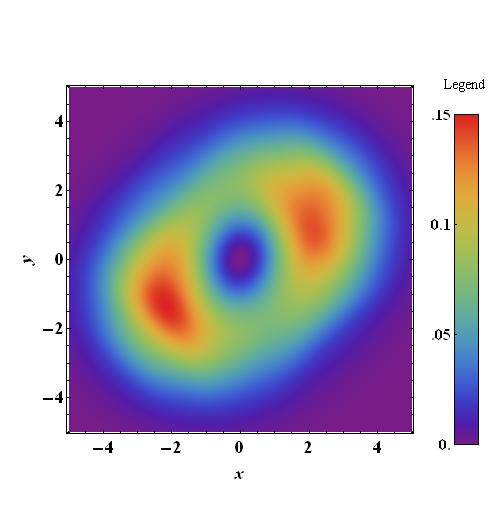}
\label{fig:label-f}}
\qquad
\subfloat[]{
\includegraphics[scale=0.2]{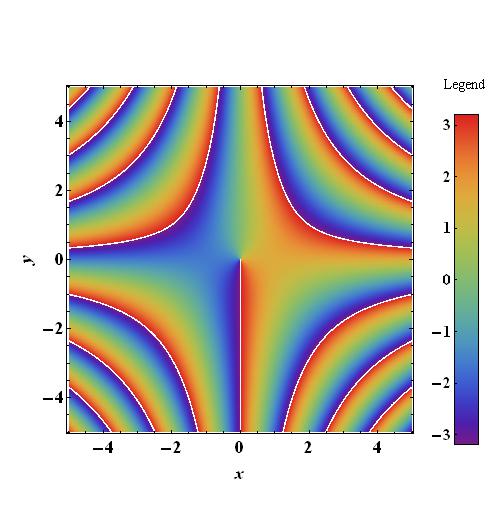}
\label{fig:label-g}}
\qquad
\subfloat[]{
\includegraphics[scale=0.2]{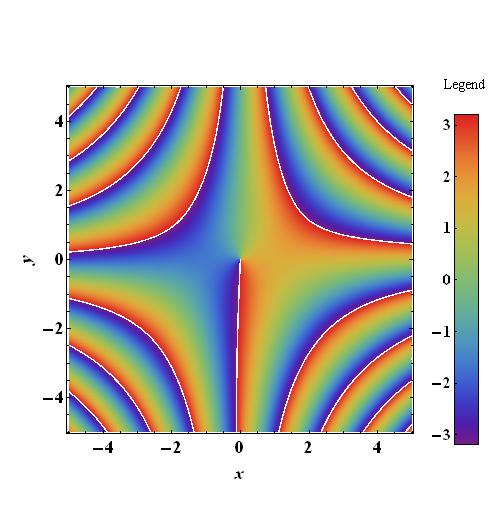}
\label{fig:label-h}}
\caption{(Color online) Contour and Phase of the state $\vert\xi\rangle^{t}_k$ (k=1) at $t=0$s (\ref{fig:label-e} and \ref{fig:label-g}) and at $t=10^{-6}$s (\ref{fig:label-f} and \ref{fig:label-h}) respectively. We have used $C=2\times 10^{10}s^{-1}$, $r=2.1$ and $\phi=\pi/2$.}
\label{fig:PSVortex}
\end{figure*}
\begin{figure*}
\centering
\subfloat[$W(x, y)_{p_y=0}^{p_x=0}$]{
\includegraphics[scale=0.2]{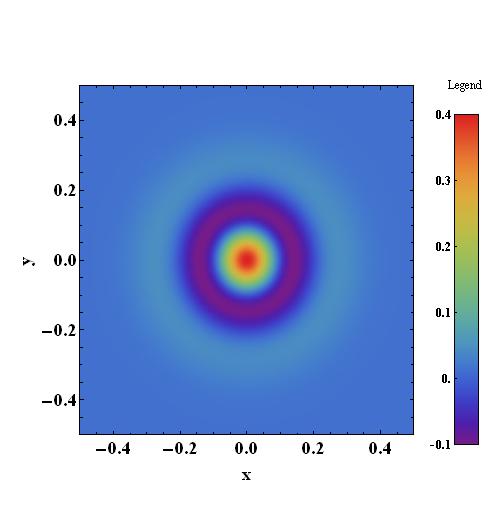}
\label{fig:label-i}}
\qquad
\subfloat[$W(x, y)_{p_x=0}^{p_y=0}$]{
\includegraphics[scale=0.2]{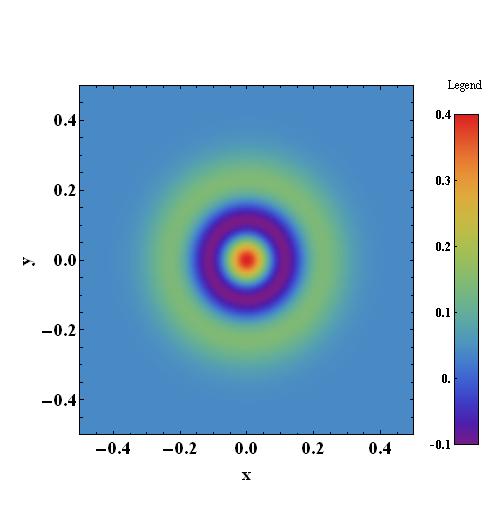}
\label{fig:label-j}}
\qquad
\subfloat[$W(y, p_x)_{p_y=0}^{x=0}$]{
\includegraphics[scale=0.2]{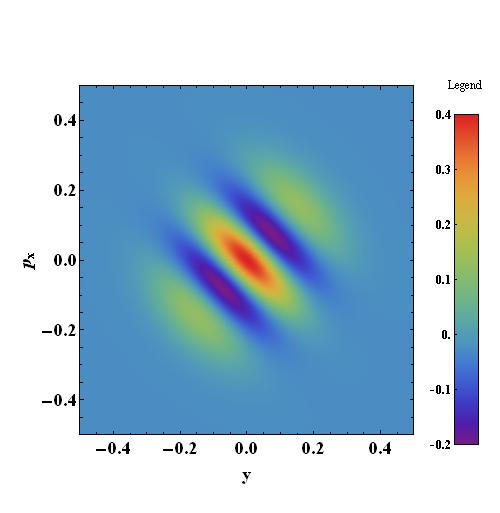}
\label{fig:label-k}}
\qquad
\subfloat[$W(y, p_x)_{p_y=0}^{x=0}$]{
\includegraphics[scale=0.2]{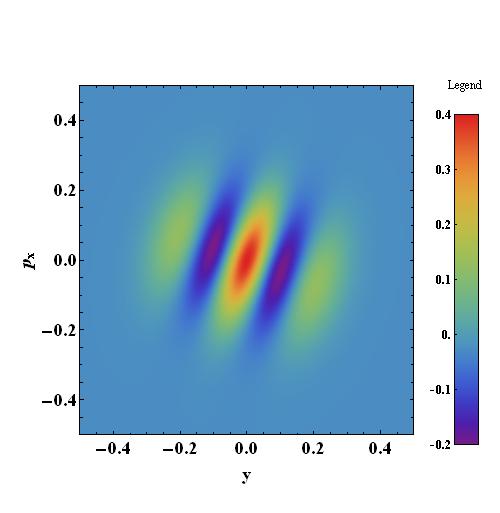}
\label{fig:label-l}}
\caption{(Color online) Contour plots of the Wigner function associated with the state $\vert\xi\rangle^{t}_k$ (k=1) at $t=0$s (\ref{fig:label-i} and \ref{fig:label-k}) and at $t=10^{-6}$s (\ref{fig:label-j} and \ref{fig:label-l}) respectively. The values of the constants are the same as in Fig. \ref{fig:PSVortex}}
\label{fig:PSVortexWigner}
\end{figure*}
In this section we study the effect of propagation through coupled waveguides on photon subtracted two mode squeezed vacuum states. It would be worthwhile to mention that these states also possess orbital angular momentum and their vortex nature is evident in the quadrature distributions \citep{gsa_njp,anindyaarxiv}. It would also be interesting to study how the vortex structure is affected due to the propagation.\\
A two mode squeezed vacuum state can be written as,
\begin{equation}
\label{TMS}
\vert \xi \rangle=\text{exp}\left(\xi a^{\dagger}b^{\dagger}-\xi^{*}ab\right)\vert 0,0\rangle, \hspace{0.2 cm} \xi=r\text{e}^{i\phi}
\end{equation}
\noindent where $\xi$ is a complex parameter, \emph{r} is the squeezing amplitude and \emph{a} and \emph{b} are the regular bosonic mode operators. If \emph{k} photons are subtracted from one of the modes, Eq. \ref{TMS} can be simplified to
\begin{eqnarray}
\label{finalstate}
\vert\xi\rangle_k &=&\frac{e^{ik\phi}}{\text{cosh}^{2}r}\sum_{m=0}^{\infty}
e^{im\phi}(\text{tanh}^mr\sqrt{m+k}\vert m+k,m\rangle \nonumber\\
&=& \frac{e^{ik\phi}}{\text{cosh}^2r} a^{\dagger k}\vert\xi\rangle
\end{eqnarray}
\noindent The state at time \emph{t} can be found out by operating it with the time evolution operator $\mathcal{U}(t)$ and solving the equation in the Schrodinger picture.
\begin{eqnarray}
\label{photonsub}
\vert\xi\rangle^{t}_k &=&\mathcal{U}(t)\vert\xi\rangle_k \nonumber \\
&=& \exp{\left[-\frac{\text{i}Ht}{\hbar}\right]}\vert\xi\rangle_k
\end{eqnarray}
\noindent where the Hamiltonian \emph{H} is same as defined in Eq. \ref{Hamiltonian}. 
We present time evolution of the state defined by Eq. \ref{finalstate} in Fig. \ref{fig:PSVortex}. The rotation produced is evident from the contour plot of the intensity. There is also a visible distortion. The order remains constant which means the orbital angular momentum is conserved.\\
The Wigner function associated with Eq. \ref{finalstate} is written as,
\begin{equation}
\label{PSWigner}
W(\tilde{\alpha}, \tilde{\beta})=\frac{4}{\pi^2}\left(-1\right)^k\mathcal{L}_k
\left[4\vert\tilde{\alpha}\vert^2\right]\exp\left[-2\left(\vert\tilde{\alpha}\vert^2 + \vert\tilde{\beta}\vert^2\right)\right]
\end{equation}
\noindent where $\mathcal{L}_k$ is the Laguerre polynomial of order \emph{k}, corresponding to the number of photons subtracted. $\tilde{\alpha}$ and $\tilde{\beta}$ are related to the coherent state parameters $\alpha = x - ip_x$ and $\beta = y - ip_y$ by a simple transformation given by,
\begin{equation}
\begin{pmatrix} \tilde{\alpha} \\ \tilde{\beta}^*\end{pmatrix} =
\begin{pmatrix}\text{cosh}r && -\text{sinh}r~e^{i\phi}\\ -\text{sinh}r~e^{i\phi} && \text{cosh}r \end{pmatrix} 
\begin{pmatrix} \alpha \\ \beta^* \end{pmatrix}
\end{equation}
To study the dynamics of the Wigner function analytically, one needs to solve the equation of motion for the Wigner function which is as follows
\begin{equation}
\label{Moyal}
\frac{\partial{W\left(\vec{r},\vec{p},t\right)}}{\partial{t}}=-\left\lbrace\left\lbrace W\left(\vec{r},\vec{p},t\right),H\right\rbrace\right\rbrace
\end{equation}
\noindent where $\lbrace\lbrace.,.\rbrace\rbrace$ is the Moyal bracket. But this is a difficult problem for most Hamiltonians and a perfect solution is known only for a few cases. In this article we study the time evolution of the Wigner function, Eq. \ref{PSWigner}, numerically. We follow the process outlined in \cite{Wong}. Given the Wigner function $W(\tilde{\alpha}, \tilde{\beta})$ at time $t=t_0$, we wish to obtain the same at $t=t_0+\delta t$ for very small $\delta t$. To do this, the phase space is divided into tiny cells centred at the phase space coordinate $\lbrace\vec{r}_0,\vec{p}_0\rbrace$ at $t_0$. For each of those coordinates, we calculate the coordinates $\lbrace\vec{r},\vec{p}\rbrace$ at a subsequent time $t$ with the condition that when $t\rightarrow t_0, \vec{r}\rightarrow \vec{r}_0$ and $\vec{p}\rightarrow \vec{p}_0$. 
In Fig. \ref{fig:PSVortexWigner} we compare the Wigner functions at the initial stage and after a time $t=10^{-6}$s. The quadratures \emph{x} and \emph{y} are correlated at $t=0$ since the outputs from a type II SPDC are entangled. The subtraction of photons further increases the entanglement. Hence the correlation pre exists. (Fig. \ref{fig:label-i}). The two modes are then propagated through a waveguide which further couples the two modes and hence there is an increase in the correlations between similar quadratures of the two modes, as evident from a higher number of concentric rings in Fig. \ref{fig:label-j}. There is no change in the correlation present between different quadratures of the two modes. It only undergoes a rotation when propagated through the waveguide which can be seen from Fig. \ref{fig:label-k} and Fig. \ref{fig:label-l}.\\

\subsection{Beam splitter/Dual channel directional coupler}
\label{sec:BS}
\begin{figure*}
\centering
\subfloat[]{
\includegraphics[scale=0.2]{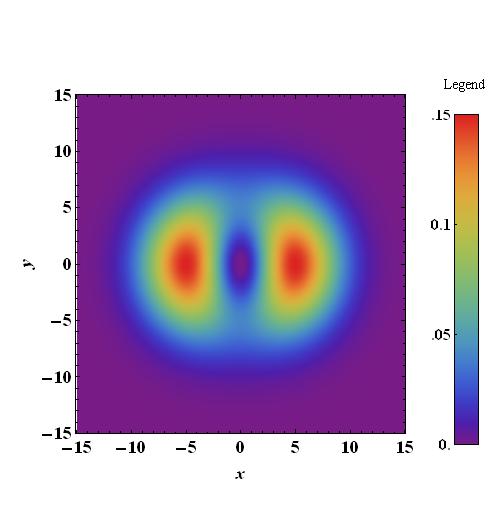}
\label{fig:label-a}}
\qquad
\subfloat[]{
\includegraphics[scale=0.2]{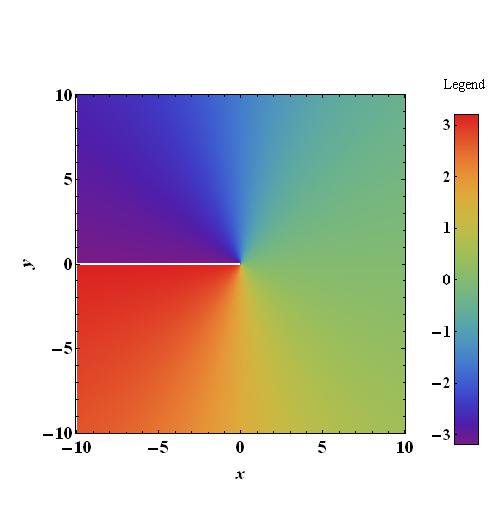}
\label{fig:label-b}}
\qquad
\subfloat[]{
\includegraphics[scale=0.2]{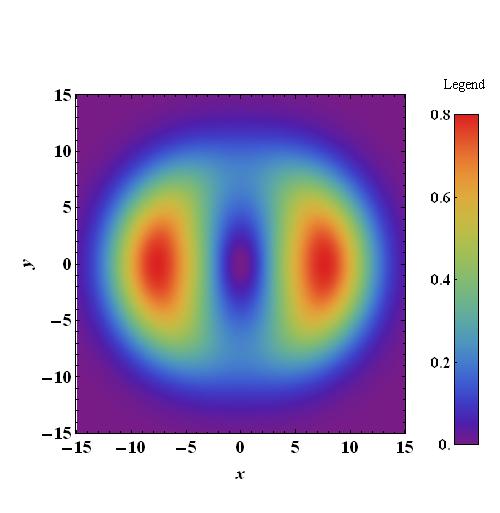}
\label{fig:label-c}}
\qquad
\subfloat[]{
\includegraphics[scale=0.2]{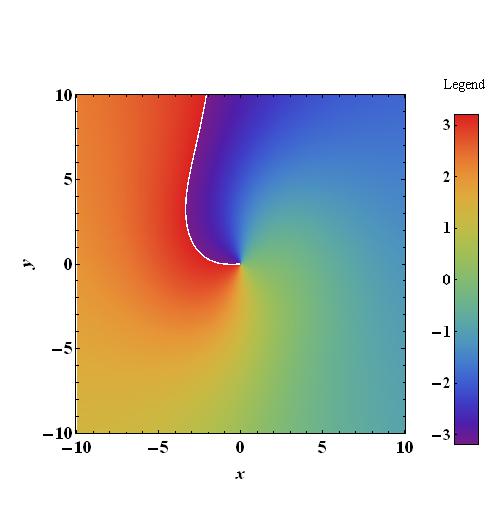}
\label{fig:label-d}}
\caption{(Color online) Contour and Phase of the state $\vert\Psi\rangle_{\text{k}}^t$ (k=1) at $t=0$s (\ref{fig:label-a} and \ref{fig:label-b}) and at $t=10^{-6}$s (\ref{fig:label-c} and \ref{fig:label-d}) respectively. We have used $C=2\times 10^{10}s^{-1}$, $r_x=0.2$, $r_y=0.5$, $\eta_x=1$ and $\eta_y=0.75$.}
\label{fig:BSVortex}
\end{figure*}
\begin{figure*}
\centering
\subfloat[$W(x,y)_{p_x=0}^{p_y=0}$]{
\includegraphics[scale=0.2]{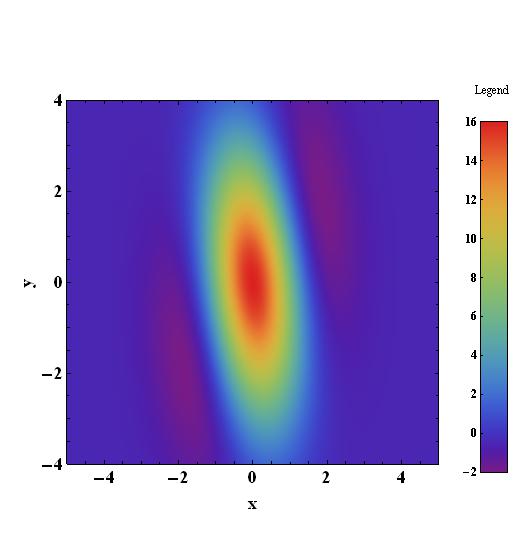}
\label{fig:label-m}}
\qquad
\subfloat[$W(x,y)_{p_x=0}^{p_y=0}$]{
\includegraphics[scale=0.2]{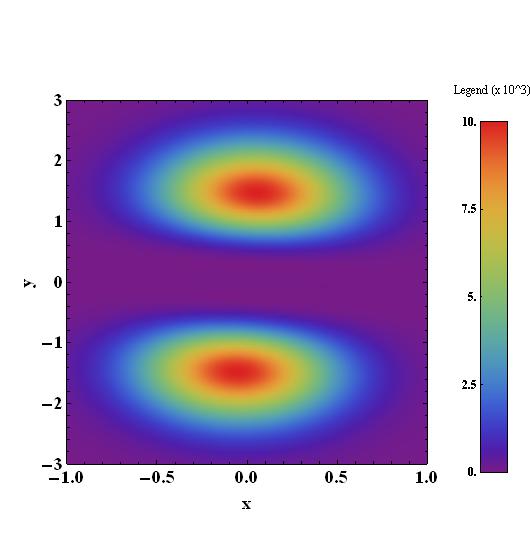}
\label{fig:label-n}}
\qquad
\subfloat[$W(y,p_x)_{x=0}^{p_y=0}$]{
\includegraphics[scale=0.2]{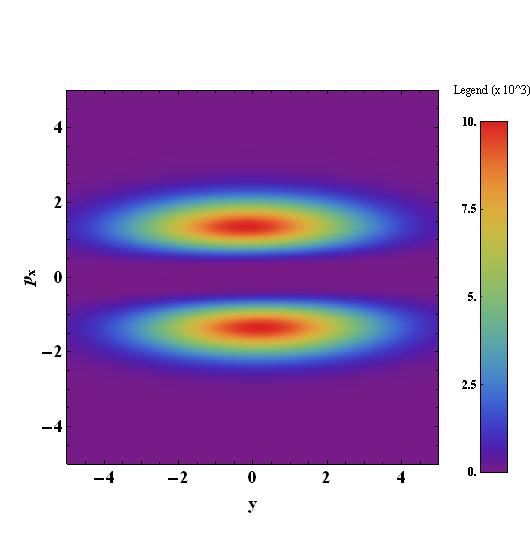}
\label{fig:label-o}}
\qquad
\subfloat[$W(y,p_x)_{x=0}^{p_y=0}$]{
\includegraphics[scale=0.2]{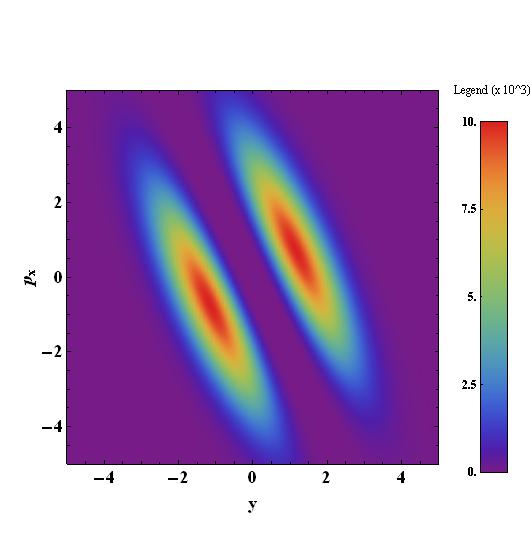}
\label{fig:label-p}}
\caption{(Color online) Contour plots of the Wigner function associated with $\vert\Psi\rangle_{\text{k}}^t$ (k=1) at $t=0$s (\ref{fig:label-m} and \ref{fig:label-o}) and at $t=10^{-6}$s (\ref{fig:label-n} and \ref{fig:label-p}). The values of the constants are the same as in Fig. \ref{fig:BSVortex}.}
\label{fig:Wigner}
\end{figure*}
Optical states carrying OAM can be generated by mixing two squeezed modes through an asymmetric beam splitter or by passing them through a dual channel directional coupler. The degree of asymmetry controls the ellipticity of the beam. A 50:50 beam splitter would give rise to a perfectly symmetrical circular vortex state. Any other ratio of mixing gives rise to an elliptical vortex which is far more general. It has also been pointed out that elliptical vortex states have higher entropy and hence higher information carrying capability \cite{ABanerji}. The generalized vortex state can be written as
\begin{eqnarray}
\label{vortex}
\vert\Psi\rangle_{\text{k}}=A\left[\eta_xa^{\dagger}-\text{i}\eta_yb^{\dagger}\right]^k\vert a,b\rangle;\\
\label{AnormConst}
\text{with}~~A=\frac{2^{-k/2}\left(1+\xi\right)^k}{\sqrt{k!}\left(\sigma_x^2+\sigma_y^2\right)^{k/2}};~\xi=\tanh{2r}
\end{eqnarray}
\\\noindent where $r=\left(r_x^2+r_y^2\right)^{1/2}$ and $r_x$, $r_y$ are the squeezing amplitudes for the individual modes, $a^{\dagger}$ and $b^{\dagger}$ are the bosonic creation operators, $\eta_x$ and $\eta_y$ are mixing parameters and $\sigma_i=\exp\left(2r_i\right)$. $\vert a,b\rangle$ is the product state of squeezed modes and can be written as,
\begin{widetext}
\begin{equation}
\label{tsmv}
\vert a,b\rangle=\frac{1}{\sqrt{\cosh r}}\sum_{m,n}\left(-1\right)^{m+n}\frac{(2m)!(2n)!}{2^{m+n}m!n!}\text{e}^{\text{i}(m+n)\phi}\tanh^{m+n}r\vert 2m,2n\rangle
\end{equation}
\end{widetext}
The quadrature distribution is obtained from Eq. \ref{vortex} by replacing the bosonic mode operators with the corresponding quadrature operators and is expressed as follows 
\begin{eqnarray}
\label{bsvortexquad}
\Psi_{\text{k}}(x,y)&=& A\left(\eta_xx-\text{i}\eta_yy\right)^k \nonumber \\
&\times &\exp\left[-\frac{1}{2}\left\lbrace\left(\frac{x}{\sigma_x}\right)^2+\left(\frac{y}{\sigma_y}\right)^2\right\rbrace\right]
\end{eqnarray}
\noindent where \emph{A} is the same normalization constant as stated in Eq. \ref{AnormConst}. The state $\vert\Psi\rangle_{\text{k}}$ after time \emph{t} can be written as
\begin{eqnarray}
\label{bsvortextime}
\vert\Psi\rangle_{\text{k}}^t &=&\mathcal{U}(t)\vert\Psi\rangle_{\text{k}} \nonumber \\
&=& \exp{\left[-\frac{\text{i}Ht}{\hbar}\right]}\vert\Psi\rangle_{\text{k}}
\end{eqnarray}
We study the time evolution of the generalized vortex state in Fig \ref{fig:BSVortex}. It is evident that the state rotates on traveling through the waveguide by comparing Fig \ref{fig:label-b} and Fig \ref{fig:label-d}. The presence of the orbital angular momentum produces the rotation. The contour also gets distorted upon propagation which can be seen by comparing Fig \ref{fig:label-a} and Fig \ref{fig:label-c}. It is interesting to compare the results of Fig. \ref{fig:PSVortex} with Fig. \ref{fig:BSVortex}. The rotation produced in the later is only in the phase whereas the former undergoes rotation in the quadrature. It would be fascinating to look at the entanglement and its behavior under propagation using the Wigner distribution function.\\
The Wigner distribution function associated with Eq. (\ref{vortex}) as shown in \cite{Abir PLA} has the following form
\begin{eqnarray}
\label{WigDisAbir}
W\left(x,y,p_{x},p_{y}\right)\hspace{4cm} \nonumber\\
= K \exp\left[-\left(X_{1}^{2}+Y_{1}^{2}+P_{X_{1}}^{2}+P_{Y_{1}}^{2}
\right)\right]\nonumber \\
\times L_{m}^{-1/2}\left[\frac{\left(P_{X_{2}}+P_{Y_{2}}-X_{2}-Y_{2}\right)^2}{\sigma_{x}^{2}+\sigma_{y}^{2}}\right]
\end{eqnarray}
\noindent where $K=\frac{2^{m-4}m!}{\pi\sqrt{\pi}\Gamma(m+\frac{1}{2})}\left[-2\left(\sigma_{x}^{2}+\sigma_{y}^{2}\right)\right]^{m}$ and $L_{m}^{-1/2}$ is the associated Laguerre polynomial. The variables in Eq. (\ref{WigDisAbir}) are a set of scaled variables defined as $X_1=\frac{x}{\sigma_x}$, $Y_1=\frac{y}{\sigma_y}$, $X_2=\frac{\sigma_yx}{2\sigma_x}$, $Y_2=\frac{\sigma_xy}{2\sigma_y}$, $P_{X_{1}}=\frac{\sigma_{x}}{\sqrt{2}}p_{x}$, $P_{Y_{1}}=\frac{\sigma_{y}}{\sqrt{2}}p_{y}$,
$P_{X_{2}}=\frac{\sigma_{y}^{3}}{\sqrt{2}}p_{x}$ and $P_{Y_{2}}=\frac{\sigma_{x}^{3}}{\sqrt{2}}p_{y}$. $\sigma_x$ and $\sigma_y$ are the standard deviations in the values of \emph{x} and \emph{y}. We follow an approach similar to that of the previous section to study the Wigner function of the generalized vortex state. In Fig. \ref{fig:Wigner} we study the Wigner function at different times. The Wigner functions at the two instances in time are seen to differ by a finite rotation. The OAM present results in producing this rotation. But there are other differences also. A nonlinear nature of the Wigner distribution is generally interpreted as arising due to the presence of correlations between the quadratures \cite{Webpage}. At time $t=0$, the state is entangled and hence there exists correlation between the two modes. At a later time, we see the correlations still exist but there is a change in their nature. There are two distinct regions where the correlations exist separated by a region of very little or no correlation as can be seen from Fig. \ref{fig:label-n}.\\
The later sections of this article deal with the entanglement between the two modes of the generalized vortex states quantitatively. We should also mention here that we have only taken into account lossless propagation. It would be interesting to see how it changes in the presence of loss, which we plan to take up later.\\

\section{Entanglement of vortex states under propagation}
\label{sec:Ent}
\begin{figure}[h!]
\includegraphics[scale=0.5]{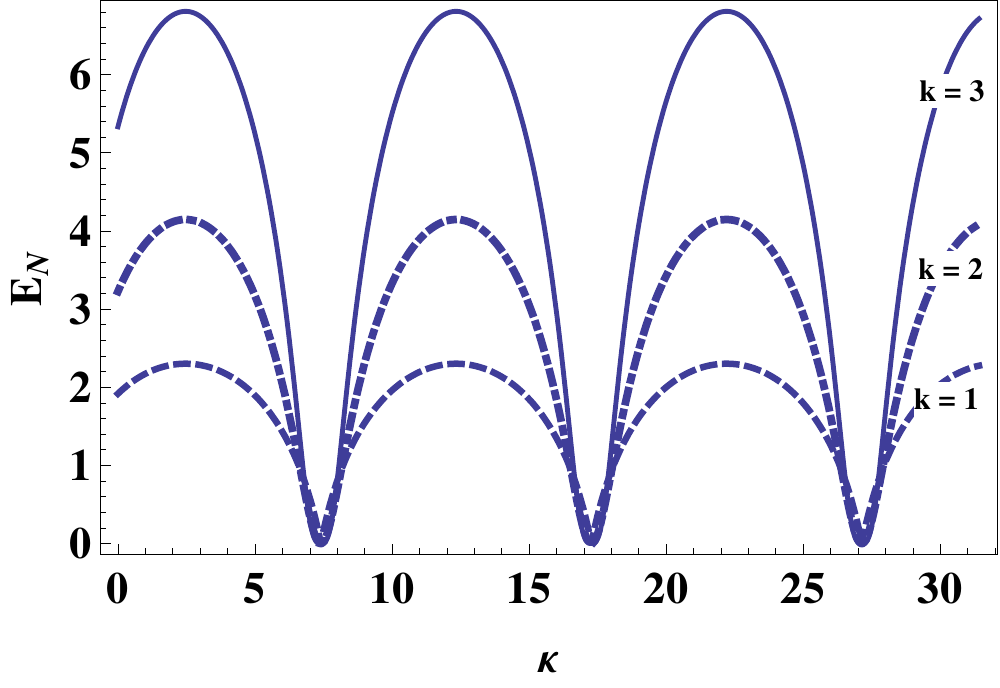}
\caption{(Color Online) \emph{Logarithmic negativity}, $E_N$ for the state $\vert\psi\rangle_{\text{k}}^t$ as a function of $\kappa=Ct/\pi$, a dimensionless quantity for different orders of the vortex. Here we have used $r_x=r_y=1$. The values of the rest of the constants are same as in Fig. \ref{fig:BSVortex}.}
\label{fig:bslognev}
\end{figure}

In this section we study the entanglement of vortex states under propagation using \emph{logarithmic negativity}. We first consider the state generated by subtracting photons from one of the modes of the output of SPDC. We use the Heisenberg picture to study the entanglement. Therefore, the mode operator in Eq. \ref{finalstate} is replaced by its time dependent counterpart. So Eq. \ref{photonsub} can be written as
\begin{eqnarray}
\label{finalstatet}
\vert\xi\rangle_k^t&=& \frac{e^{ik\phi}}{\text{cosh}^2r} a^{\dagger k}(t)\vert\xi\rangle
\end{eqnarray}
\noindent where $a^{\dagger}(t)$ is the same as described in Eq. \ref{modeoperatorstimedep}. Now Eq. \ref{finalstatet} can be written as
\begin{widetext}
\begin{equation}
\label{PSVortexT}
\vert\xi\rangle_k ^t=\frac{e^{ik\phi}}{\text{cosh}^2r}\sum_n \text{e}^{in\phi}\tanh^n r\sum _m ^k \left(\begin{array}{ccc}k\\m\end{array}\right)\cos^{k-m}(Ct)\sin^m(Ct)i^m\sqrt{\frac{(n+k-m)!(n+m)!}{n!^2}}\vert n+k-m,n+m\rangle
\end{equation}
\end{widetext}
\noindent The corresponding density matrix is constructed from Eq. \ref{PSVortexT}. It has the following form
\begin{widetext}
\begin{eqnarray}
\label{PSVdensity}
\rho=\sum_{n,q}\frac{\tanh^{n+q}r}{\cosh^4r}\sum_{m,p}c_mc_p(-i)^{m+p}\vert n+k-m,n+m\rangle\langle q+k-m,q+m\vert\text{e}^{i(n-q)\phi} \\
\text{where}~~~c_m=\left(\begin{array}{ccc}k\\m\end{array}\right)\cos^{k-m}(Ct)\sin^m(Ct)\sqrt{\frac{(n+k-m)!(n+m)!}{n!^2}}
\end{eqnarray}
\end{widetext}
\noindent The partial transpose of the density matrix can be easily obtained from Eq. \ref{PSVdensity}. Since a density matrix is hermitian and a hermitian matrix cannot have imaginary eigenvalues, we only consider those terms for which $m=p$. Taking this into account we see that all the eigenvalues for even \emph{m} are negative. We use these eigenvalues to determine the \emph{logarithmic negativity}, which is defined as follows
\begin{equation}
\label{lognev}
E_N(\rho)=\log_2\left(1+2\mathcal{N}(\rho)\right)
\end{equation}
\noindent where $N$ is the sum of all the negative eigenvalues.\\
\begin{figure*}
\centering
\subfloat[$k=2$]{
\includegraphics[scale=0.5]{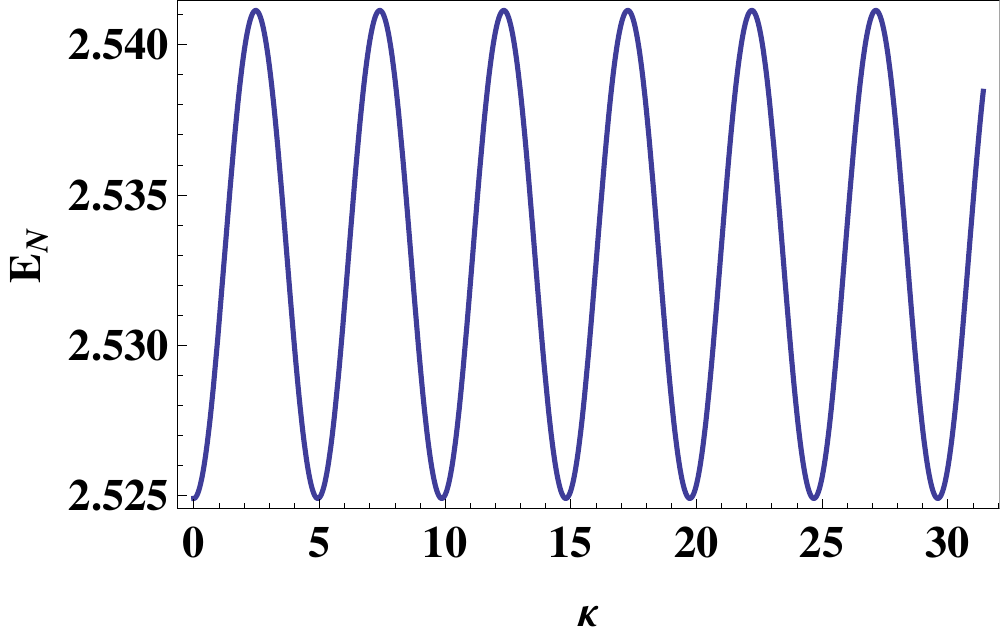}
\label{k2}}
\qquad
\subfloat[$k=3$]{
\includegraphics[scale=0.5]{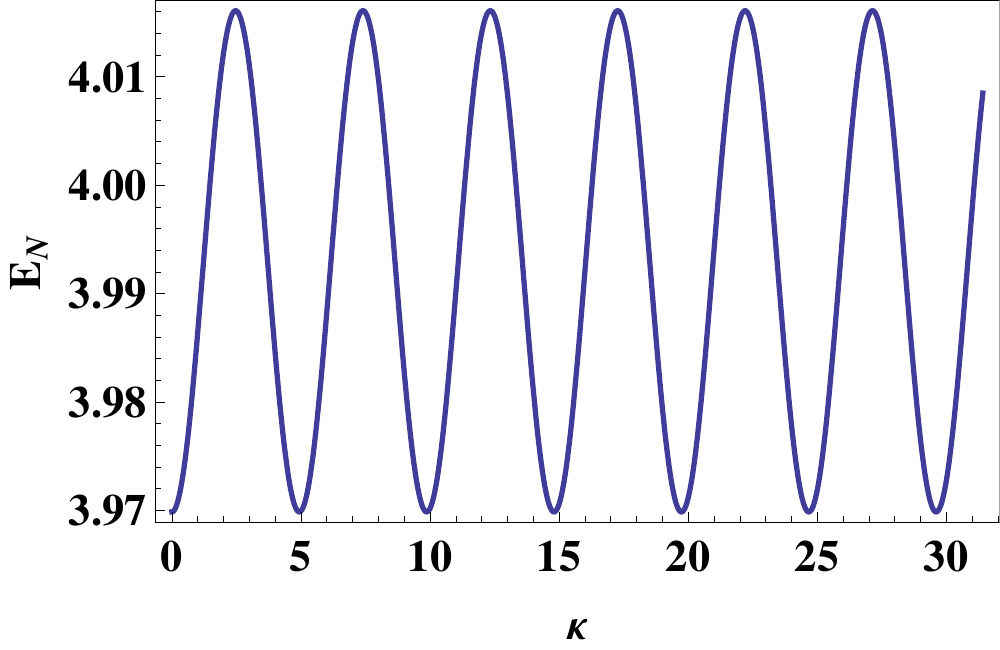}
\label{k3}}
\caption{(Color online)\emph{Logarithmic negativity}, $E_N$ for the state $\vert\xi\rangle_k ^t$ as a function of $\kappa=Ct/\pi$, a dimensionless quantity, for different orders of the vortex. Here we have used $C=2\times 10^{10}s^{-1}$ and $r=1$.}
\label{fig:LogNevPSV}
\end{figure*}
In Fig. \ref{fig:LogNevPSV} we study the variation of $E_N$ with time for different orders of the vortex state $\vert\xi\rangle_k^t$. We observe that $E_N$ oscillates between a maximum and a minimum value, the difference between which is only around 0.6$\%$ of the initial value for $k=1$. The input state being already entangled, $E_N=2.525$ at $t=0$ for $k=2$. On entering the coupled waveguides, the two modes get further entangled due to the coupling between the two modes of the waveguide and $E_N$ increases to 2.541 at $\kappa=2.5$. It again reduces to its original value at $\kappa=5$. This same behavior is periodically repeated for later times. A higher order vortex state being more entangled compared to a lower order one, we observe that the initial value and the maximum value reached for a vortex state of higher order is larger. However, the periodic behavior as well as the period of oscillation remains unchanged for higher orders as can be seen from Fig \ref{k3}.\\
Next we take a look at the vortex state described by Eq. \ref{vortex}. In a similar approach we replace the mode operators with their time dependent counterparts. Eq. \ref{vortex} at time \emph{t} can then be written as
\begin{widetext}
\begin{equation}
\label{bsvortexT}
\vert\psi\rangle_{\text{k}}^t=\frac{A}{\cosh r}\sum_{m,n}(-1)^{m+n}\frac{\sqrt{(2m)!(2n)!}}{2^{m+n}m!n!}\left[\eta_xa^{\dagger}(t)-\text{i}\eta_yb^{\dagger}(t)\right]^k\text{e}^{\text{i}(m+n)\phi}\tanh^{m+n}r\vert 2m,2n\rangle
\end{equation}
\end{widetext}
Using Eq. \ref{modeoperatorstimedep} to substitute $a^{\dagger}(t)$ and $b^{\dagger}(t)$, we can write Eq. \ref{bsvortexT} as,
\begin{widetext}
\begin{equation}
\label{bsvortexT2}
\vert\psi\rangle_{\text{k}}^t=\frac{A}{\cosh r}\sum_{m,n}(-1)^{m+n}\frac{\sqrt{(2m)!(2n)!}}{2^{m+n}m!n!}\left[\left(\eta_xa^{\dagger}-\text{i}\eta_yb^{\dagger}\right)\cos \left(Ct\right)+\left(\eta_ya^{\dagger}+\text{i}\eta_xb^{\dagger}\right)\sin \left(Ct\right)\right]^k\text{e}^{\text{i}(m+n)\phi}\tanh^{m+n}r\vert 2m,2n\rangle
\end{equation}
\end{widetext}
\noindent The corresponding density matrix constructed from Eq. \ref{bsvortexT2} has the following form
\begin{widetext}
\begin{eqnarray}
\label{bsvortexDensity}
\rho=N\sum_{m,n,q,p}\sum_{j,l}^k c^{mn}_j c^{qp}_l\vert 2m+k-j,2n+j\rangle ~ \langle 2q+k-l,2p+l\vert\text{e}^{\text{i}(m+n-q-p)\phi}\\
\text{where}~~c^{mn}_j=(-1)^{m+n}\left(-\text{i}\right)^j\left(\begin{array}{ccc}k\\j\end{array}\right)\frac{\tanh^{m+n}r}{2^{m+n}m!n!}\sqrt{(2m+k-j)!(2n+j)!}{cs_+}^{k-j}{cs_-}^j;\\
c^{qp}_l=(-1)^{q+p}\left(\text{i}\right)^l\left(\begin{array}{ccc}k\\l\end{array}\right)\frac{\tanh^{q+p}r}{2^{q+p}q!p!}\sqrt{(2q+k-l)!(2p+l)!}{cs_+}^{k-l}{cs_-}^l;\\
cs_{+}=\eta_x\cos\left(Ct\right)+\eta_y\sin\left(Ct\right);
~cs_{-}=\eta_y\cos\left(Ct\right)-\eta_x\sin\left(Ct\right)\nonumber
\end{eqnarray}
\end{widetext}
\noindent where $N=A^2/\cosh^2r$. We take a partial transpose to determine all the negative eigenvalues of this density matrix. We consider only those cases for which $j=l$. This helps us to get rid of all the imaginary eigenvalues and we are thus left with only real numbers. We determine $E_N$ defined in Eq. \ref{lognev} for this case and study it in Fig. \ref{fig:bslognev}.\\
It is seen that $E_N$, in this case also, oscillates between a maximum and minimum value with time. The entanglement between the two modes increases with increasing time, reaches a maximum and then falls off to zero and starts rising again. The input state being an entangled state, $E_N=1.8$ at $t=0$. It increases to 2.4 at $\kappa=2.25$. It then goes to zero at $\kappa=7.25$ at which stage the two modes of the vortex state are fully separable. This behavior continues for the total duration with a specific periodicity. A higher order of the vortex only increases the maximum value reached by $E_N$ but no change in the oscillating behavior is observed. This is because the oscillations arise due to the propagation through the coupled waveguide and the initial state preparation plays no part in this.
\section{Conclusion}
\label{sec:Conc}
In this article we have studied the time evolution of generalized quantum optical vortex states. We have compared the results for two of the more popular methods of generation of these states i.e. one by the method of photon subtraction/addition from one of the modes of the output of a SPDC and another by using a series of beam splitters to couple two squeezed modes. We studied the propagation of these states through coupled lossless waveguides using the Schrodinger picture. We observed that due to the presence of orbital angular momentum, the states have gone through rotation under propagation. But there was a difference between the behavior of two states. Although the order remained unchanged in both the cases, the photon subtracted vortex state underwent a rotation in the quadrature whereas the other type of vortex state underwent a rotation only in the phase while spreading out in the quadrature distribution. We constructed the Wigner function at a later time for both type of states. It was seen, apart from a rotation visible in the cross - correlation between different quadratures, the states propagated with no loss in quality. This is because we have considered lossless waveguides as a model. Hence it will be interesting to see how it behaves in the presence of loss which we plan to deal with in a future work. There were also some oscillations visible when we reconstructed the Wigner function for the state $\vert\psi\rangle_{\text{k}}^t$. These oscillations can be attributed to the coupling present between the two modes of the waveguide.\\
We have also explicitly calculated the entanglement present in these states using \emph{logarithmic negativity}. We observed oscillatory behavior for both the states when propagated through the waveguide. Since both the states were entangled prior to the start of propagation, $E_N$ had a non - zero value at $t=0$. It changed periodically from an entangled state to a fully separable state for $\vert\psi\rangle_{\text{k}}^t$. As the states become separable, the coupling between the two modes of the waveguide results in correlating the two modes of the state which increases entanglement until it reaches the maximum and falls off again. The photon subtracted vortex state, $\vert\xi\rangle_k ^t$, on the other hand exhibited contrasting behavior. Although the periodic change in entanglement was visible under careful examination, the entanglement did not vanish completely. It was found to oscillate between the initial value of entanglement and a maximum value. This behavior is in sharp contrast to a two mode squeezed state. The difference arises mainly because of the non - Gaussian nature that is added to the state by subtracting photons which results in entanglement distillation.
\acknowledgments{
This work is partially sponsored by DST through SERB grant no. SR/S2/LOP - 0002/2011.}

\end{document}